\documentstyle[epsfig,mncite]{mn}
\ifCUPmtlplainloaded \else

\fi
\title[Optical and X-ray Variability in Cyg X-2]{High-speed Keck II and RXTE spectroscopy of Cygnus X-2: (I) Three X-ray components revealed by correlated variability}

\author[K. O'Brien et al.]{K. O'Brien$^{1,2,3}$, Keith Horne$^2$, Richard H.
Gomer$^4$, J. B. Oke$^{5,6}$, M. van der Klis$^{3}$\\
        $^1$ European Southern Observatory, Alonso de Cordova 3107, Santiago, Chile \\
        $^2$ School of Physics and Astronomy, University of St. Andrews, St. Andrews KY16 9SS \\
        $^3$ Astronomical Institute ``Anton Pannekoek'', University of Amsterdam, 1098-SJ Amsterdam, The Netherlands \\
        $^4$ Howard Hughes Medical Institute and Department of Biochemistry and
Cell Biology, MS-140, Rice University, Houston, \\ TX 77005-1892, USA \\
        $^5$ California Institute of Technology, Mail Stop 105-24, Pasadena, CA 91125, USA\\ 
	$^6$ Dominion Astrophysical Observatory, Herzberg Institute of Astrophysics, National Research Council of Canada, \\ 5071 West Saanich Road, Victoria, BC V8X 4M6, Canada \\
}

\date{Accepted ... . Received ...; in original form 2000 ....}

\pagerange{\pageref{firstpage}--\pageref{lastpage}}

\begin{document}
\label{firstpage}
%
% Define stellar names
%
\newcommand{\cyg}{Cygnus~X-2}
\newcommand{\novasco}{GRO\,J1655--40}
\newcommand{\novaper}{GRO\,J0422+32}
\newcommand{\novamus}{X-ray Nova Muscae 1991}
\newcommand{\sco}{Scorpius X-1}
\newcommand{\bhb}{Black Hole Binary }
%
% Define telescope names
%
\newcommand{\HST} {\textit{HST}}
\newcommand{\xte} {\textit{RXTE}}
\newcommand{\rxte}{\textit{RXTE}}
\newcommand{\GRO} {\textit{GRO}}
\newcommand{\keck}{\textit{Keck II}}
%
% Define names of ions
%
\newcommand{\HI}   {H\,\textsc{i}}
\newcommand{\HII}  {H\,\textsc{ii}}
\newcommand{\HeI}  {He\,\textsc{i}}
\newcommand{\HeII} {He\,\textsc{ii}}
\newcommand{\HeIII}{He\,\textsc{iii}}
\newcommand{\halpha}{H\,$\alpha$} 
\newcommand{\hbeta}{H\,$\beta$} 
\newcommand{\hgamma}{H\,$\gamma$} 
%
% Define reddening symbols
%
\newcommand{\EBV}{E(B-V)}
\newcommand{\Rv} {R_{\rm V}}
\newcommand{\Av} {A_{\rm V}}
%
% Define wavelength symbols
%
\newcommand{\lam}   {$\lambda$}
\newcommand{\lamlam}{$\lambda\lambda$}
\newcommand{\mdot}{\dot{M}}
\newcommand{\msolar}{M_{\sun}}
\maketitle
%
% Define command to produce comment
%
\newcommand{\comm}[1]{\textit{[#1]}}
\newcommand{\object}{Cygnus X-2}
\newcommand{\utstart}{12:52}
\newcommand{\utend}{13:06}
\newcommand{\etime}{72.075}
\newcommand{\skyfrac}{0.9}
\newcommand{\nspec}{220,000}
\newcommand{\binfac}{14}
\newcommand{\bintime}{1.01}
\newcommand{\nbin}{1646}
\newcommand{\degree}{$^{\circ}$}
\newcommand{\rank}{$S_z$}
%
%%%%%%%%%%%%%%%%%%%%%%%%%%%%%%%%%%%%%%%%%%%%%%%%%%%%%%%%%%%%%%%%%%%%%%%%%%%%%%%
%
\begin{abstract}
We have performed simultaneous X-ray and optical spectroscopic observations of the Low Mass X-ray Binary \cyg. We have used a new data system attached to the Low Resolution Imaging Spectrograph instrument on Keck II to obtain spectra with a mean time resolution of \etime\ milliseconds, simultaneous with pointed X-ray observations using the PCA onboard \xte. In this paper, we have analysed the variability in both wavebands on timescales of 16 seconds. During our observations \cyg\ covered all three branches of the Z-curve, allowing us to study how the changes in X-ray spectral state affect the optical emission. As the optical flux rises the X-ray intensity first rises on the Horizontal Branch ($0<S_z<1$) but then falls on the Normal Branch ($1<S_z<2$) and Flaring Branch ($2<S_z<3$), where \rank\ is a rank number characterising the position on the Z-curve. This linear increase in the optical flux with \rank\ indicates the optical flux is a good predictor of the accretion rate (possibly normalized by its own long-term average) inferred from the Z-state $S_z$. We have used this correlation to decompose the total X-ray count-rate into three distinct spectral components. 
\end{abstract}
%
%%%%%%%%%%%%%%%%%%%%%%%%%%%%%%%%%%%%%%%%%%%%%%%%%%%%%%%%%%%%%%%%%%%%%%%%%%%%%%%
%
%
%%%%%%%%%%%%%%%%%%%%%%%%%%%%%%%%%%%%%%%%%%%%%%%%%%%%%%%%%%%%%%%%%%%%%%%%%%%%%%%
%
\section{Introduction}
\cyg\ is one of the best studied low-mass X-ray binaries (LMXBs) due to its X-ray variability and observable optical counterpart. \cyg\ is a close binary comprising a 1.78\,$\msolar$ neutron star in a 9.84 day binary orbit with a 0.6 $\msolar$ sub-giant companion star \cite{orosz99}. Since its discovery in 1965 \cite{bowyer65} there have been a host of observations in the optical (eg. \scite{paradijs90}), ultraviolet (eg. \scite{vrtilek90}) and X-ray (eg. \scite{kuulkers96}) wavebands. The optical counterpart of \cyg, V1341 Cygni, has a spectral type A9III and contributes about 50\% of the total visual flux \cite{casares98}. The reprocessing of X-rays in \cyg\ is shown by the large contribution of the disc to the UV continuum and emission line fluxes, which arise predominantly from the X-ray heated accretion disc \cite{dejong96}.

\cyg\ belongs to the sub-class of LMXBs known as Z-sources, whose X-ray spectral states trace a Z-shaped pattern in X-ray colour-colour diagrams (see \pcite{hasinger89}). In a similar manner to the colour-colour diagrams familiar to optical astronomers, whose colour indices are magnitude differences between two bands, X-ray colour-colour diagrams employ ``hardness ratios'', which are ratios of counts in a high-energy band to counts in a low-energy band. Typically a hard colour (hardness ratio for two high-energy bands) is plotted against a soft colour (hardness ratio for to low-energy bands). The X-ray equivalent of a colour-magnitude diagram is a hardness-intensity diagram, which plots a hardness ratio against the X-ray intensity (count~s$^{-1}$) in some band. The location of the source in these diagrams provides information about changes in the shape of the X-ray spectrum and intensity, which arise for example from changes in the mass accretion rate onto the neutron star. In the case of the Z-sources, they seem to move smoothly along three distinct tracks, from the Horizontal Branch, the upper-most stroke, through the Normal Branch, the diagonal stroke, and onto the Flaring Branch, the lowest stroke, a sequence that until recently has been interpreted as a monotonic increase in the mass accretion rate from below to above the Eddington limit \cite{hasinger89}. There is however mounting evidence that this simple relationship is not the whole story and that a further parameter is needed to fully describe the X-ray behaviour. This has been demonstrated most noticeably with the ``parallel lines'' phenomena, where the relationship between the X-ray luminosity and QPO frequency is valid on short timescales only, but changes on the duration of days to weeks, indicating a second parameter is needed to fully characterise the X-ray emission of LMXBs (see \scite{vdk2001} for a discussion of this phenomenon).

\begin{figure}
\begin{center}
\epsfig{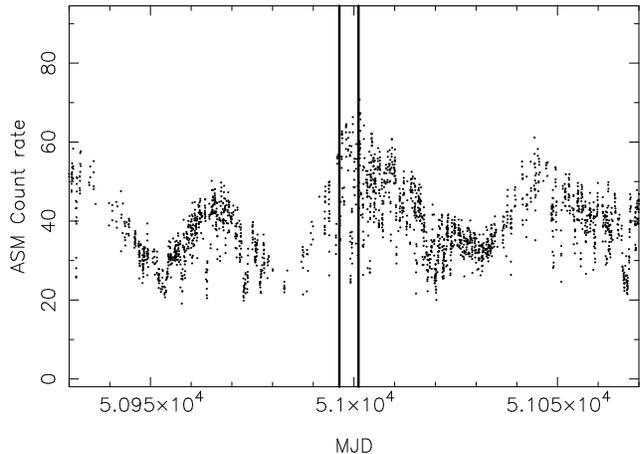}
\caption[ASM lightcurve for \cyg]{The RXTE All Sky Monitor lightcurve for Cygnus X-2. The vertical lines show the beginning and end of the Keck and RXTE observations used in our analysis.}
\label{cygasmplot}
\end{center}
\end{figure}

The position of the Z in the X-ray colour-colour diagram and the shapes of the soft and hard colour-intensity diagrams evolve on a timescale of weeks \cite{kuulkers96}. During this time \cyg\ varies between three levels; a high, medium and low level, as defined by \scite{kuulkers96}. The low intensity level is rare and only observed in the binary phase range 0.8-0.2, indicating it may be linked to a grazing eclipse of the X-ray emitting region by the mass donor star. More normally, the system appears to move slowly between the medium and high intensity levels. This can be seen in the \rxte\ ASM lightcurve, a section of which is shown in Figure~\ref{cygasmplot}. The source can be seen to move slowly between the medium level, with a count rate of $\sim$ 30 counts/sec and the high level where the count rate is $\sim$ 50 coutns/sec. However, large, short timescale deviations from these levels can also be seen. These sudden drops in the count rate are caused by motion along the Z-curve and are not thought to be related to the longer timescale changes (the so-called ``secular variations''). The upper edge of this envelope represents the count rate at the Hard Vertex, between the Horizontal and Normal Branches, where the count rate is highest. 

Both the colour-colour diagrams and colour-intensity diagrams for the medium and high intensity levels are very different. In the high intensity level, the hard colour intensity diagram shows linear, almost horizontal Horizontal and Flaring Branches, whilst in the medium level these branches are more curved. The Flaring Branch is also truncated during the High level, which has been interpreted as obscuration by material in the binary, whereas in the medium level, the Flaring Branch intensity varies smoothly with the X-ray intensity. It is thought that the truncation is most noticeable during the Flaring Branch as it coincides with the highest mass accretion rate, which causes the inner disc to swell and shield our view of the X-ray emitting regions (see \scite{kuulkers96} for a more detailed description). 

\cyg\ exhibits Type I X-ray bursts (eg. \scite{kuulkers95}), confirming the presence of a neutron star, as seen in many other neutron star binaries, including the Z-source GX\,17+2 (\scite{Kahn84}, \scite{kuulkers97b}). These bursts typically have rise times of 0.1-10 seconds and are thought to be due to thermonuclear burning of accreted material on the surface of the neutron star \cite{smale98}. Unlike neutron stars, black holes have no stellar surface and Type I X-ray bursts are not observed. There is also tentative evidence for a third periodicity in \cyg\ with a period of 78 days, although the interpretation of this is complicated by the changes in X-ray flux as \cyg\ moves around the colour-colour diagram. The long-term quasi-periodicity can be seen in Figure~\ref{cygasmplot}, where the peak-to-peak separation is $\sim30-40d$. \scite{wijnands96} associate this periodicity with that of a precessing accretion disk, as seen in Hercules~X-1 \cite{giacconi73}.

A previous multi-wavelength campaign on \cyg\ during June and October 1988, involving X-ray \cite{hasinger90}, UV \cite{vrtilek90}, optical \cite{paradijs90} and radio \cite{hjellming90} observations, revealed many interesting features of the correlated variability, which are summarized schematically in Figure 9 of \scite{hasinger90}. They found that there was a direct correlation between the UV and optical fluxes and the position on the Z-diagram, although the sampling of the optical and UV data ($\sim$ 100's of seconds) only allowed general trends to be observed. They also found a more complicated behaviour between the X-ray intensity and the position on the Z diagram. In this series of papers we will build upon these observations, with higher time resolution optical data ($\sim$ seconds), which enables us to observe the previously observed trends in much more detail.

In this paper, the first in the series, we shall discuss the observations of the X-ray and optical spectral variability of \cyg\ in terms of broadband variability on timescales of $10^{2-5}$ seconds. We will also go into some depth into the data reduction tasks performed, which will be abridged in subsequent papers. In subsequent papers, we will discuss the variability of narrow band features in the optical spectra and how this is affected by the X-ray spectral state. We will also discuss a simultaneous timing analysis of the data and the nature of the rapid ($\sim$ seconds) correlated X-ray and optical variability in terms of the time delayed optical emission relative to the driving X-ray variability. 

%
%%%%%%%%%%%%%%%%%%%%%%%%%%%%%%%%%%%%%%%%%%%%%%%%%%%%%%%%%%%%%%%%%%%%%%%%%%%%%%%
%
\section{Data}
Simultaneous Keck and RXTE observations of \cyg\ were taken during UT 1998 July 2-6 as part of our campaign to search for correlated optical and X-ray variability in X-ray binaries. In this paper, we will show the optical data from the first and last nights of our campaign, where atmospheric conditions were most favourable. An analysis of the entire X-ray dataset showed that using only this subset did not reduce our coverage of the different spectral states of \cyg.
%
%%%%%%%%%%%%%%%%%%%%%%%%%%%%%%%%%%%%%%%%%%%%%%%%%%%%%%%%%%%%%%%%%%%%%%%%%%%%%%%
%
\subsection{Optical}
The optical data were taken using the Low Resolution Imaging Spectrograph (LRIS; \scite{oke95}) on the 10-m Keck II telescope on Mauna Kea, Hawaii. A summary of the observations is given in Table~\ref{files}. The LRIS was used with a 5.2 arcsecond slit masked with aluminized mylar tape to form a square aperture.  The 300/5000 grating used has a mean dispersion of 2.55 \AA/pixel in the range 3600\AA\ - 9200\AA. 

We used a novel data acquisition system, which reads out the CCD continuously, to obtain more than \nspec\ spectra of \cyg, in the form of continuous byte streams lasting typically 50 minutes. The mean integration time was measured to be \etime\-ms and there is no dead-time between individual spectra. 

In addition to the 2048-pixel spectrum, 25-pixel under-scan and 75-pixel over-scan regions were used to measure the CCD bias level, which we subtracted from each spectrum. The noise for a given pixel was calculated using a readout noise of 6.3 e$^-$ and a gain of 4.7 e$^-$/ADU. Cosmic rays were rejected with a threshold of 10 sigma from the de-biased frames. A master flatfield image was created by finding the median of 700 individual flatfield spectra. This image showed no deviations above 0.3 \% in all but 3 pixels. It was decided that it was therefore not necessary to flatfield individual spectra. Calibration arc spectra and spectra of the sky in the region of the object were taken at regular intervals. 

The background spectrum, which accounts for $\sim$ \skyfrac\% of the total flux, conains two components; one from scattered light within the spectrograph and a second from the sky brightness. The scattered light level was measured from the counts in the pixels below $3400\AA$, where both the GRISM and atmospheric transmission are negligable. This level was removed from all spectra and then the sky spectrum was subtracted using the best fit to spectra taken of a blank sky field. These sky spectra were taken at the beginning and end of the exposure, so that long timescale variations could be detected. The mean and variable components of these spectra were found by creating a lightcurve for each pixel and extracting the mean and gradient of this lightcurve. In order to supress the pixel-to-pixel variations, the mean and gradient were then averaged in wavelength, with a running median filter of width 101 pixels. The gradient in time of the data was found to be consistent with zero, indicating that the spectrum was constant for the duration of the exposure ($\sim$ 50~minutes). This spectrum is more than a simple sky spectrum and contains both the spectrum of the sky backgound and an additional component thought to be due to scattered light within the spectrograph. 

The wavelength calibration was accomplished by fitting a second-order polynomial to 7 lines in a median spectrum of exposures of Hg and Ar arc lamps. Arc spectra from the beginning and end of the run were used to take into account any drifts in the wavelength scale. The wavelength calibration was applied using the MOLLY spectral analysis package. 

\begin{figure}
\begin{center}
\epsfig{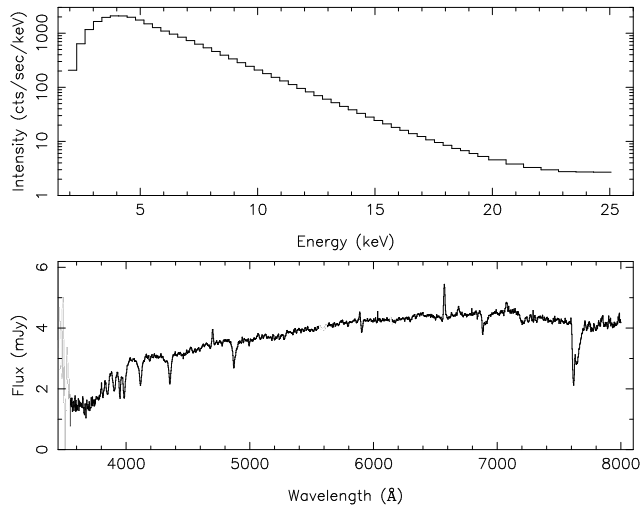}
\caption[Typical spectra]{The average X-ray (top panel) and optical (bottom panel) spectra from observation 'G'. The X-ray spectrum is plotted with a logaritmic scale.}
\label{specsplot}
\end{center}
\end{figure}

The individual spectra of \cyg\ were flux calibrated using exposures taken of the standard star Feige 67 \cite{oke90}. We fitted a low-order polynomial to the median spectrum of the observations of this standard star. The resulting fit was used as a flux calibration and again applied using MOLLY. The average spectrum from observation 'G' is shown in Figure~\ref{specsplot}. Several emission and aborption features can be cleary seen and will be described in detail in Paper~II.

%

%
%%%%%%%%%%%%%%%%%%%%%%%%%%%%%%%%%%%%%%%%%%%%%%%%%%%%%%%%%%%%%%%%%%%%%%%%%%%%%%%
%
\subsubsection{Time calibration}
Due to the nature of the continuous stream of data, it was impossible to attach accurate timemarks to each individual spectrum. Individual time marks were placed after every other spectrum using the computer clock. In order to establish an absolute time reference, secondary timestamps were obtained on many occasions during the 5-night observing run. To create these timestamps, an incandescent lamp that illuminated the CCD was modulated in synch with WWVH time signals. From these timestamps we found an ephemeris for the observation and the mean exposure time. These were checked against the computer's time marks and found to be in excellent agreement. 

The input times were corrected to the solar system barycentre, using the subroutines supplied by the International Astronomical Union's (IAU) working group on ``Standards Of Fundamental Astronomy'' (SOFA\footnote{http://www.iau-sofa.rl.ac.uk/}). Finally, these times were converted to seconds since JD 2450996.0, in order to make the numbers more manageable.

In order to increase the signal-to-noise of the individual spectra and make the data-set more manageable, the spectra were binned in time by a factor of \binfac\ to give \nbin\ binned spectra with a time resolution of \bintime\ seconds.

\begin{table*}
\begin{center}
\begin{tabular}{|c|c|c|c|c|c|c|}
Filename & Observation & Night & XTE window & UT at start & Elapsed time & Number of unbinned \\
& & & & & (secs) & spectra \\ \hline
K0010 & A & 1 & 1 & 10:48, 02/07/98 & 1654 & 23268 \\
K0013 & B & 1 & 2 & 13:15, 02/07/98 & 3930 & 58086 \\
K4005 & C & 5 & 1 & 09:08, 06/07/98 & 1460 & 20230 \\
K4007 & D,E & 5 & 2,3 & 10:05, 06/07/98 & 3890 & 46942 \\
K4009 & F & 5 & 4 & 11:42, 06/07/98 & 3746 & 51926 \\
K4011 & G & 5 & 5 & 13:17, 06/07/98 & 1710 & 23842 \\
\end{tabular}
\end{center}
\caption[Keck II observations summary]{Summary of the Keck II optical observations. The binning factor of 14 used in the lightcurves gives an integration time of 1.01 seconds for the resulting spectra.}
\label{files}
\end{table*}
%

%
%%%%%%%%%%%%%%%%%%%%%%%%%%%%%%%%%%%%%%%%%%%%%%%%%%%%%%%%%%%%%%%%%%%%%%%%%%%%%%%
%
\subsection{X-ray}
The X-ray data taken with the PCA onboard the \rxte\ satellite covered 42 ksec of exposure time during UT 1998 July 2-6. We have used a subset of this data, which amounts to 20.3 ksec and represents the times of simultaneous coverage during favorable atmospheric conditions. During this time all 5 PCUs were switched on. The data were analysed using the \textsc{ftools v5.1} software suite. The \textsc{Standard-2} mode data were used with a time resolution of 16 seconds to create colour-colour and hardness-intensity diagrams, and to study the simultaneous variability. The lightcurves were extracted, using \textsc{ftools} package \textsc{saextrct}, in 4 energy ranges, using the \textsc{Standard-2} channels 1-5, 6-12, 13-21 and 22-39. This corresponds to 1.94 - 3.72 keV, 3.72 - 6.22 keV, 6.22 - 9.46 keV and 9.46 - 16.02 keV. The times were corrected to the solar system barycentre using the \textsc{ftools} package \textsc{faxbary}. The minimum count rate of 4000 count~s$^{-1}$ implies that background subtraction was not necessary. The deadtime is estimated to be $<\,2\%$, which is smaller than the variability we are studying and therefore a deadtime correction was not applied to the X-ray lightcurves. The average spectrum during observations 'G' is shown in Figure~\ref{specsplot}.

The broadband X-ray spectral variations of Z-sources are most clearly shown in an X-ray colour-colour diagram. These are shown for Cygnus X-2 in, e.g.,  \scite{wijnands01}. The colour-colour and hardness intensity diagrams presented by them contain data from the 5 nights of observations. In this paper we present a subset of this data for which we have simultaneous optical observations. The colour-colour diagram from our \rxte\ observation is shown in Figure~\ref{allccdiagram}. The soft colour is defined as the intensity in the range 3.72 - 6.22 keV, divided by the intensity in the range 1.94 - 3.72 keV. The hard colour is similarly defined as the intensity in the range 9.46 - 16.02 keV divided by the intensity in the range 6.22 - 9.46 keV.

%
%%%%%%%%%%%%%%%%%%%%%%%%%%%%%%%%%%%%%%%%%%%%%%%%%%%%%%%%%%%%%%%%%%%%%%%%%%%%%
%
\section{The broadband X-ray spectral changes}

\begin{figure*}
\begin{center}
\epsfig{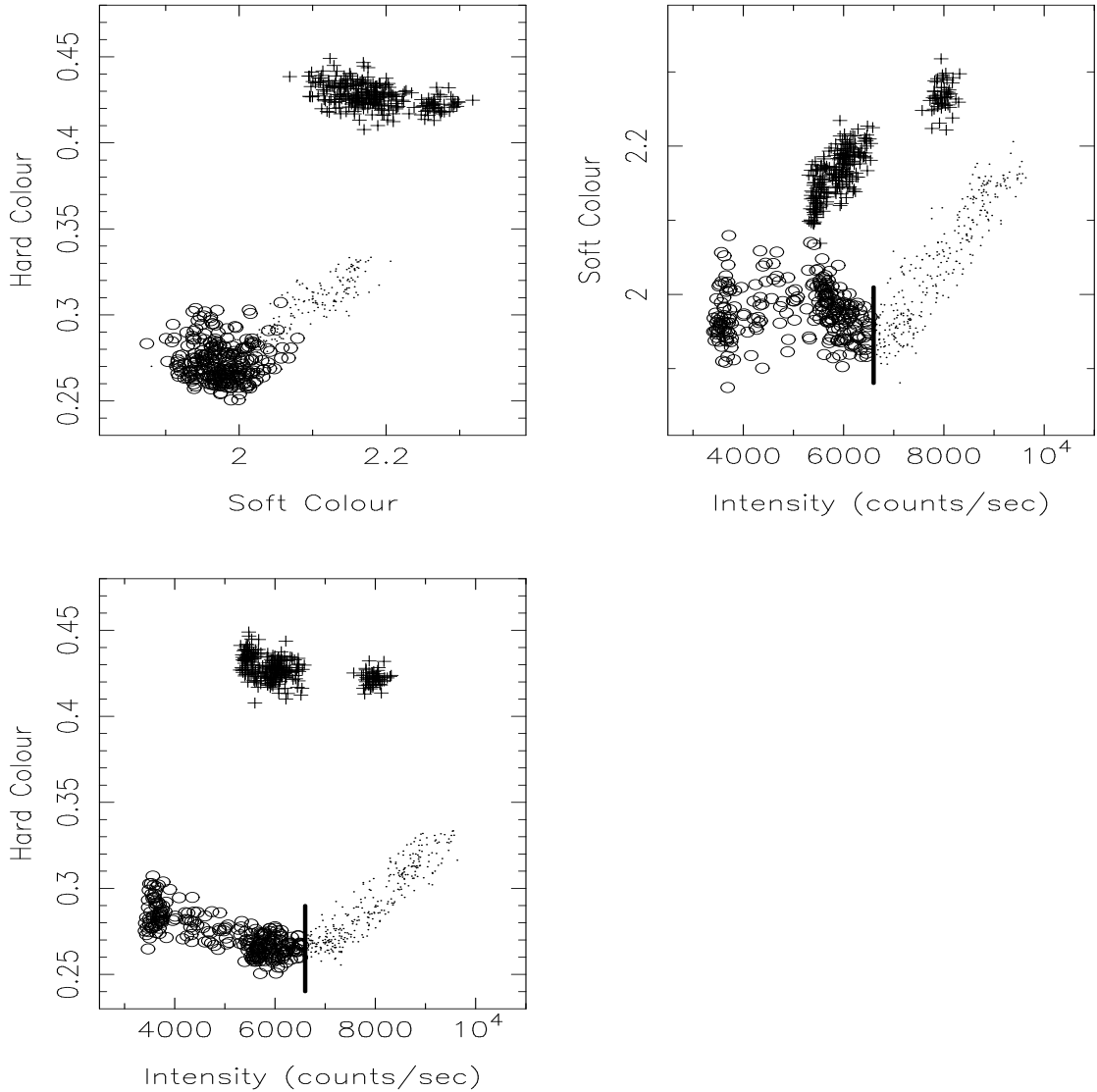}
\caption[]{Top left, X-ray colour-colour diagram for Cygnus X-2, based on the \xte\ PCA data between UT 01/07/98 - 06/07/98. The data are taken from the \textsc{Standard-2} mode, with a time resolution of 16 seconds. The soft colour is defined as the ratio of the intensities in the ranges 1.94 - 3.72 keV and 3.72 - 6.22 keV. The hard colour is the ratio of the intensities in the ranges 6.22 - 9.46 keV and 9.46 - 16.02 keV (an increase in colour means the spectrum hardens). Bottom left, the hardness intensity diagram for the hard colour and the total intensity. Top right, the hardness intensity diagram for the soft colour and the total intensity. The Horizontal, Normal and Flaring Branches are indicated by pluses, dots and open circles respectively. The vertical line shows the X-ray intensity at the Normal/Flaring Branch vertex.}
\label{allccdiagram}
\end{center}
\end{figure*}

\cyg\ can clearly be seen in Figure~\ref{allccdiagram} to trace out all three branches of the Z-curve, beginning on the Horizontal Branch during the first night, then moving onto the lower Normal Branch and the Flaring Branch during the last night. If motion along the `Z' is due to changes in the mass accretion rate (see \scite{hasinger90}), we would infer that $\mdot$ increases from night to night.

The hardness intensity diagrams, which plot the X-ray colour changes as a function of intensity show that the hard colour varies smoothly with intensity whilst on a given branch, as expected from previous observations of the source. The transition between the Normal Branch and the Flaring Branch, referred to as the soft vertex, is clearly seen, but the Horizontal Branch to Normal Branch transition region, referred to as the hard vertex, is not covered in any of our observations. 

The soft colour-intensity diagram also shows a linear trend on the Horizontal and Normal Branches. This is interpreted as changes in the X-ray production mechanism as a function of the increasing mass accretion rate. The soft colour-intensity diagram deviates from the expected linear relationship (see eg. \scite{kuulkers96}) whilst on the Flaring Branch. The colour-intensity diagrams are characteristic of the high intensity level colour-intensity diagram described by \scite{kuulkers96}, exhibiting an almost horizontal FB in the hard colour-intensity diagram.

%

%
%%%%%%%%%%%%%%%%%%%%%%%%%%%%%%%%%%%%%%%%%%%%%%%%%%%%%%%%%%%%%%%%%%%%%%%%%%%%%%%
%
\section{The simultaneous X-ray and optical variability in Cygnus X-2}
\label{cyglightssection}
\subsection{General features of Night 1, UT 01-02/07/98}

\begin{figure*}
\begin{center}
\epsfig{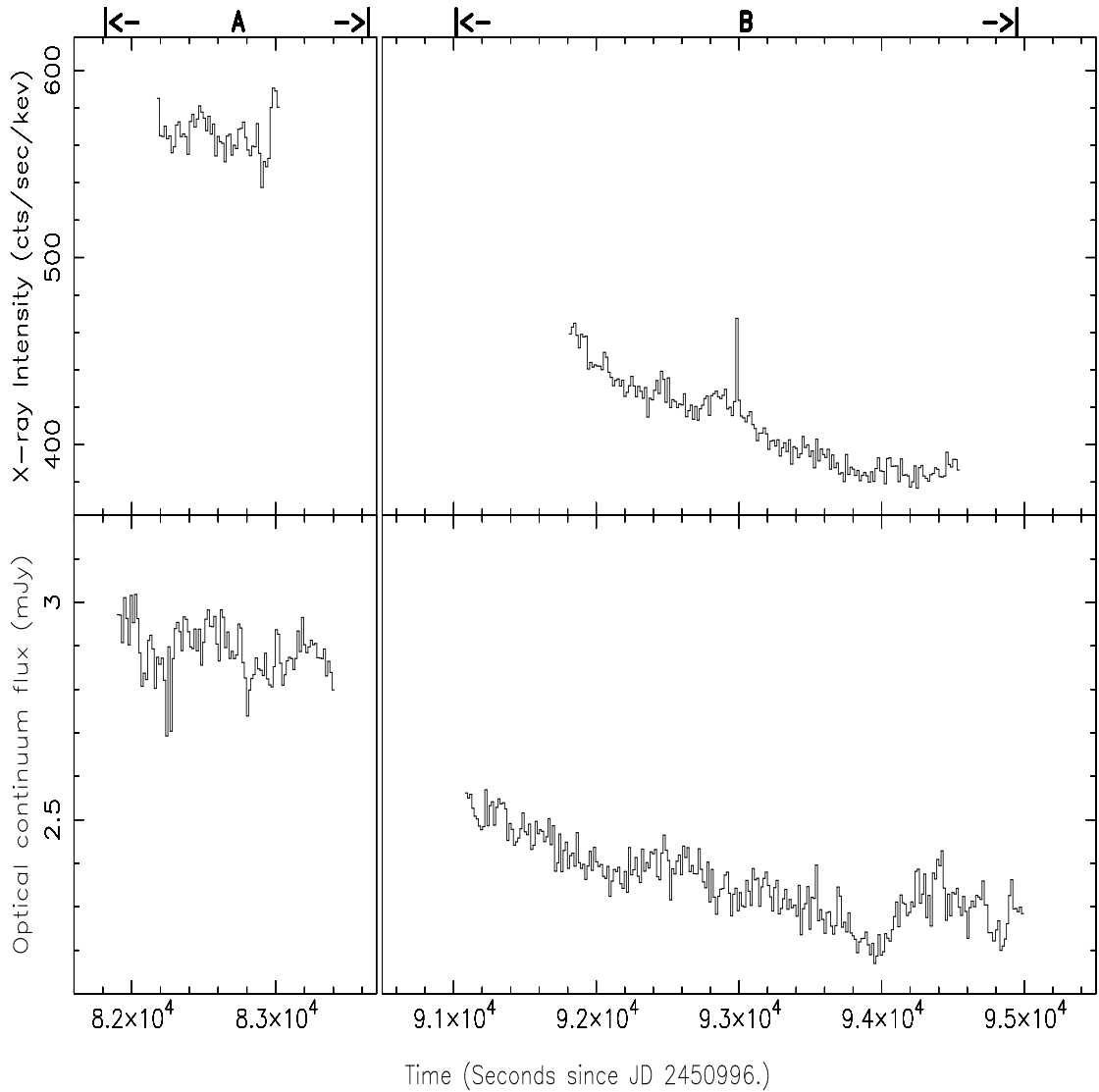}
\caption[X-ray and optical lightcurves from 01/07/98]{Top panels, the total X-ray intensity during the 2 RXTE visits (labeled A,B) from night 1. Bottom panels, the optical continuum flux in the range 5000 - 5800 $\AA$. The time is seconds elapsed since JD 2450096.0, the beginning of the night. The time resolution of the lightcurves is 16 seconds. The relative sizes of the left and right panels has been chosen to keep the scale the same for both lightcurves. The X-ray burst can clearly be seen at $t=93,000$~secs, with no optical counterpart.}
\label{cygintplot1}
\end{center}
\end{figure*}

The simultaneous X-ray and optical observations of \cyg\ from Night 1 of our observations are shown in Figure~\ref{cygintplot1}. The position in the X-ray colour-colour diagram (the pluses in Figure~\ref{allccdiagram}) show that \cyg\ was on the Horizontal Branch throughout the night. It began the night near the Hard Vertex, but by the end of the night had moved along the Horizontal Branch away from the Hard Vertex, which coincides with the lowest X-ray and optical fluxes. During these observations there is a general correlation between the X-ray and optical flux. In contrast to the Normal and Flaring Branch lightcurves shown in Figure~\ref{cygintplot} and described in the next section, no rapid ($10^{\,0-2}$ seconds) correlated variability is seen on the Horizontal Branch. Even the X-ray burst observed around $t=93,000$~secs shows no optical counterpart. (Correlated, time-delayed variability on these timescales will be analysed in a subsequent paper in this series.)

\subsection{General features of Night 5, UT 05-06/07/98}

\begin{figure*}
\begin{center}
\epsfig{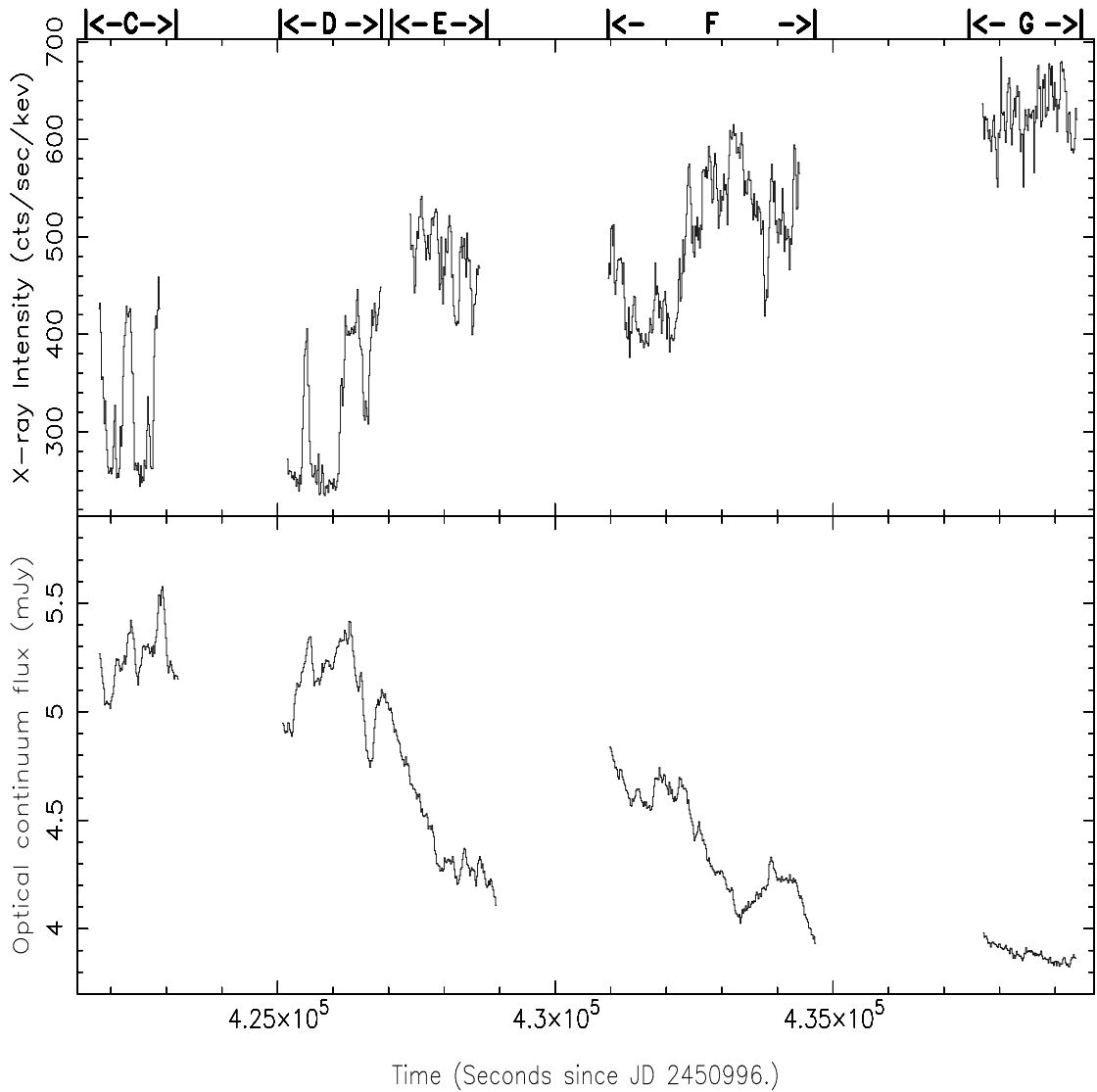}
\caption[X-ray and optical lightcurves from 06/07/98]{Top panel, the total X-ray intensity during the 5 RXTE visits (labeled C-G) from night 5. Bottom panel, the optical continuum flux in the range 5000 - 5800 $\AA$. The time is seconds elapsed since JD 2450996.0, the beginning of the first night of the run. The time resolution of the lightcurves is 16-seconds.}
\label{cygintplot}
\end{center}
\end{figure*}

At the beginning of Night 5, \cyg\ was on the Flaring Branch. During the night it moved off the Flaring and onto the Normal Branch, finishing approximately half-way between the Soft and Hard Vertices. The overall X-ray intensity is now anti-correlated with the optical flux, as shown in Figure~\ref{cygintplot}. Although this might seem surprising, it is in fact the same behaviour as reported by \scite{hasinger90} between the X-ray intensity and UV flux. Their interpretation for this was that the reprocessing material in the disk, which gives rise to much of the observed optical/UV emission, intercepts X-rays from a number of different lines of sight, whereas our observations are limited to the line of sight from the neutron star to the detector. As these lines of sight may have quite different fluxes, averaging over many of them gives a better indication of the bolometric X-ray flux and hence $\mdot$. They note that the observed X-ray flux is sensitive to both the changing spectrum of the source and any anisotropy in the X-ray emission. Perhaps the existence of anisotropic emission could also explain the lack of an optical counterpart to the X-ray burst, as the burst is seen by the observer, but not the region of the binary responsible for reprocessing the X-ray photons into optical/UV photons. 
%

%
%%%%%%%%%%%%%%%%%%%%%%%%%%%%%%%%%%%%%%%%%%%%%%%%%%%%%%%%%%%%%%%%%%%%%%%%%%%%%%%
%
\section{The correlations between the X-ray and optical data}
\label{cygcorsection}
\begin{figure}
\begin{center}
\epsfig{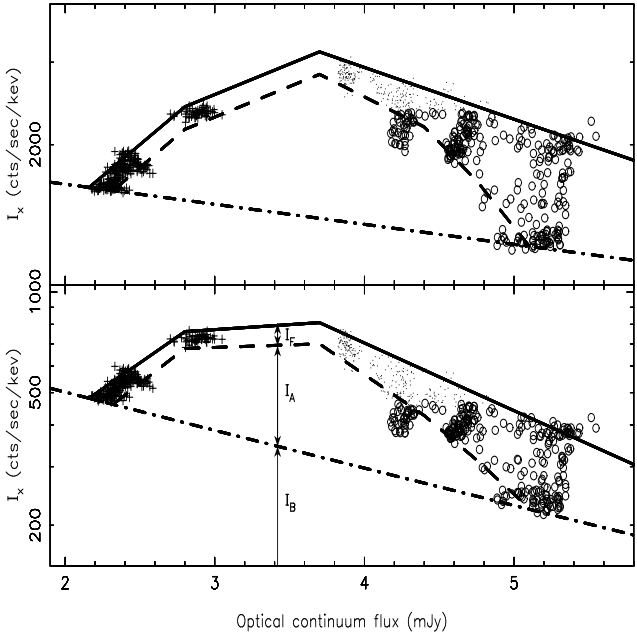}
\caption[X-ray and optical flux variability in \cyg]{Top, a plot of the logarithm of the Soft X-ray intensity (1.94\,-\,6.22~keV) versus optical continuum flux. Bottom, the same plot for the Hard X-ray intensity (6.22\,-\,16.02~keV). The solid, dotted and dot-dashed lines represent the limits of the three components described in the text. The Normal, Horizontal and Flaring Branch points are represented by pluses, dots and open circles respectively.}
\label{cygrelplot}
\end{center}
\end{figure}
In Figure~\ref{cygrelplot} we plot the X-ray intensity in the soft (top panel) and hard (bottom panel) bands as a function of the optical continuum flux. A striking feature is the significant range in X-ray intensity $I_X$ that occurs at each optical flux (the error bars are typically smaller than the markers in the plot).  This indicates there is not a simple one-to-one relationship between the X-ray and optical fluxes of \cyg. The range over which $I_X$ varies, at a given optical flux ($F_\nu$), remains rather small on the Horizontal Branch (shown as pluses in Figure~\ref{cygrelplot}), but it then increases dramatically as the system progresses up the Normal and Flaring Branches (marked by dots and open circles respectively). It appears plausible from Figure~\ref{cygrelplot} to define smooth upper and lower envelopes representing the extreme values of $I_X$ that occur at each optical flux. These envelopes are indicated by the solid and dashed lines on Figure~\ref{cygrelplot}. 

On the basis of the optical and X-ray variability exhibited in Figure~\ref{cygrelplot}, we investigate here a decomposition of the X-ray intensity $I_X$ into three spectral components, $I_{\rm F}$, $I_{\rm B}$ and $I_{\rm A}$, where the subscripts stand for Flaring, Baseline and Accretion respectively (see Section~\ref{compsection}). This decomposition can be expressed mathematically as 
\begin{equation}
I_{X}(F_{\nu}) =  \alpha(t)\,I_{\rm F}(F_{\nu}) + I_{\rm B}(F_{\nu}) + I_{\rm A}(F_{\nu}),\label{intensityeqn}
\end{equation}
where $\alpha(t)$ is a coefficient to describe the variability in $I_{\rm F}$, with $0\leq\alpha(t)\leq1$. The three components are discussed below and shown schematically in the lower panel of Figure~\ref{cygrelplot}.
 
\subsection{The three components}
\label{compsection}

The first component, $I_{\rm F}$, with F for ``flaring'', is associated with the rapid variation between the bounds of the envelope. The variability of this component is mimicked in the optical lightcurve, as can be seen in the correlated flares and dips during Observations C and D, shown in Figure~\ref{cygintplot}. In other words, on the Normal and Flaring Branches, the area between the full and dashed lines in Figure~\ref{cygrelplot} is filled in by the source performing diagonal (correlated) strokes on the timescale of $10^{1-2}$ seconds.

A second interesting feature in Figure~\ref{cygrelplot} is the lower envelope or ``baseline'' in the range of $I_X$ when $F_\nu > 4.8$~mJy. This baseline in X-ray intensity is visited repeatedly by the system during observations C and D in Figure~\ref{cygintplot}, where it appears to represent a level below which $I_X$ never goes. We notice in Figure~\ref{cygrelplot} that this floor in $I_X$ decreases as the optical flux rises. Furthermore, if we extrapolate that slope to lower optical flux it appears to predict the lowest $I_X$ we observed on the Horizontal Branch, at \rank$=0$. We therefore draw a line across the diagram, and identify the X-ray intensity of this baseline component as $I_{\rm B}$, B for ``baseline'', where $I_{\rm B}$ decreases linearly as the optical flux rises.

The third component is simply the difference between the baseline intensity $I_{\rm B}$ and the lower envelope of $I_X$ at each optical $F_\nu$. This component first increases along the Horizontal Branch, and then decreases along the Normal and Flaring  Branches.  We refer to the X-ray intensity of this component that varies smoothly with the optical flux as $I_{\rm A}$, ``A'' for accretion, since we suggest that it is related to the accreted material (see discussion).

Our decomposition is somewhat arbitrary. For example, we could combine $I_{\rm B}$ and $I_{\rm A}$ into a single component with a more complex dependence on optical flux. However, as described above, we believe that the morphology of the optical and X-ray variations, especially the apparent floor in the $I_X$, provides sufficient motivation to further explore our proposed 3-component decomposition.

%Our spectral decomposition also represents a departure from the usual connection between the X-ray spectral state, \rank, and accretion rate.  We identify the optical flux and component $I_{\rm A}$ with the accretion rate. $I_{\rm A}$ and $S_Z$ are practically coincident on the horizontal branch, but they exhibit distinct variations on the flaring branch where $I_{\rm A}$ varies slowly with the optical flux while \rank\ varies rapidly with $I_X$.

There are several deviations outside this envelope, which we will describe on a case-by-case basis. The most obvious of these deviations occurs around $F_\nu = 4.3$mJy, where $I_X$ is below the lower edge of the envelope. This deviation can be seen in the X-ray lightcurve in Figure~\ref{cygintplot} around $t=434,000$~secs in the form of a dip in the X-ray lightcurve that has no optical counterpart. Such ``non-correlated'' dips have been seen previously in \cyg\ \cite{kuulkers96} and interpreted as momentary obscuration by material along the line of sight to the neutron star. They are not expected to have an analogue in the optical lightcurve as the line of sight from the X-ray emitting region to the reprocessing region is different from the direct line of sight for the observer. Similar non-correlated dips are seen in the Flaring Branch observations, which result in the dips seen around $F_\nu = 4.2$mJy, where the X-ray intensity drops to the lower ``floor'' level (described in the next section). 

Another deviation, seen at the highest optical fluxes is associated with the reprocessed flares seen during Observation C in Figure~\ref{cygintplot}. While we would expect such correlated flares to remain within the envelope, as they do appear to have X-ray analogues, we have not taken into account the time delay between the X-ray and optical emission. This time delay, which will be quantified in another paper in this series, is most noticeable when the amplitude of the variability is largest on timescales similar to the time resolution of the data, as is the case on the Flaring Branch. This rapid variability can cause spurious correlations, as is the case here. 

The variability outside of the envelope at low optical fluxes $F_\nu = 2.1-2.2$mJy is not understood and more data will be needed in order to determine the cause of the non-correlated variability.

\subsection{The spectra of the components}

Comparing the upper and lower panels of Figure~\ref{cygrelplot}, showing the relationship between the X-ray and optical fluxes for the soft and hard X-ray bands respectively, we see how the 3 components depend on the X-ray band. The most striking feature of this is how the hard X-ray intensity (lower panel) reaches its maximum at lower $F_{\nu}$ than the soft X-ray intensity.

\begin{figure}
\begin{center}
\epsfig{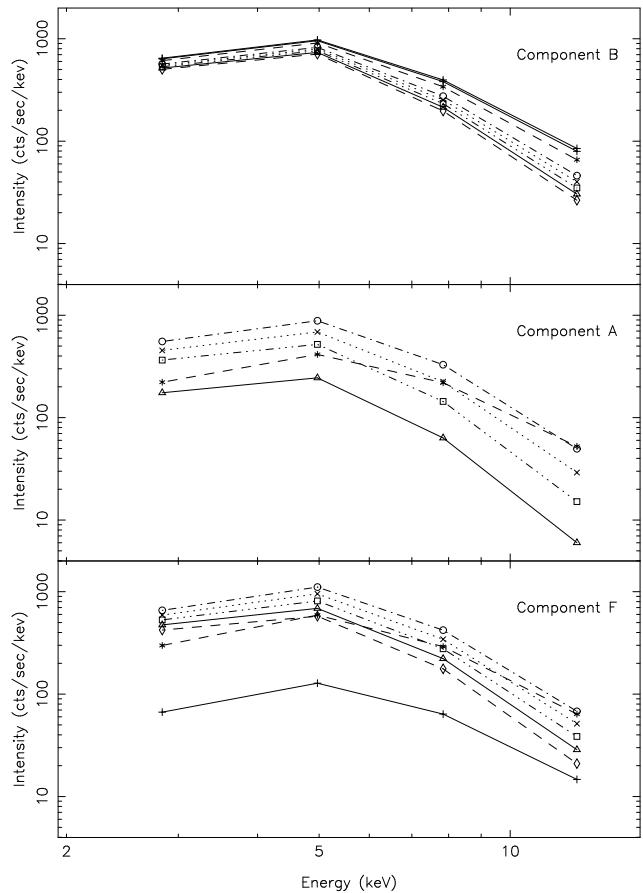}
\caption[]{The spectra of the three identified components in the X-ray spectra.The components are $I_{\rm B}$, $I_{\rm A}$ and $I_{\rm F}$ from top to bottom panel respectively. The spectra were calculated when the optical continuum flux equaled 2.3 (pluses, solid line), 2.8 (stars, dashed line), 3.7 (circles, dot-dashed line), 4.05 (crosses, dotted line), 4.4 (squares, dot-dot-dot-dashed line), 4.75 (triangle, solid line) and 5.1 (diamond, dashed line). The spectra for $I_{\rm A}$ when $F_{\nu}$ = 2.3 \& 5.1 are set to zero, as can be seen from Figure~\ref{cygrelplot}. }  
\label{cygthreespecs}
\end{center}
\end{figure}

In order to determine the spectrum of each component and how it varies with the optical flux, we have sub-divided the X-ray band into a total of four bands. These bands are the same as those chosen to create the hardness ratios shown in Figure~\ref{allccdiagram}, namely 1.94 - 3.72 keV, 3.72 - 6.22 keV, 6.22 - 9.46 keV and 9.46 - 16.02 keV. The results of this decomposition are shown in Figure~\ref{cygthreespecs}. $I_{\rm B}$ (Top Panel) is seen to have a roughly constant soft X-ray flux, while the Hard X-ray flux drops steadily, leading to an overall softening of the spectrum. This is in contrast to $I_{\rm F}$ (Bottom Panel), where the soft flux varies most. $I_{\rm A}$ shows the least spectral variability, showing only a change between the transition from the Horizontal Branch to the Normal Branch.

%

%
%%%%%%%%%%%%%%%%%%%%%%%%%%%%%%%%%%%%%%%%%%%%%%%%%%%%%%%%%%%%%%%%%%%%%%%%%%%%%%%
%
\section{Summary and Discussion}
We have obtained a unique dataset from Keck II and \xte\ and used it to investigate the simultaneous X-ray and optical variability in \cyg. We have found that on timescales, $\sim 10^{2-5}$ seconds, the X-ray and optical variability show a positive-correlation on the Horizontal Branch, changing to an anti-correlation on the Normal and Flaring Branches. 

However, on short timescales, $\sim 10^{0-2}$~seconds , the optical lightcurves show correlated, yet time-delayed features of the X-ray lightcurves. This is as expected in a linear reprocessing model. (The short timescale correlated variability will be discussed in more detail in a later paper.) The simultaneous X-ray and optical fluxes show a complicated correlation. Initially, whilst on the Horizontal Branch they show a fairly tight correlation. Then, when \cyg\ is on the Normal Branch, the X-ray and optical flux changes become anti-correlated, whilst still exhibiting a strong correlation. Then, around the middle of the Normal Branch, the correlated variability changes again: the correlation becomes weaker with large changes in both the X-ray and optical variability. 

We have used the simultaneous X-ray intensity and optical flux measurements to characterise the spectrum in terms of the three components $I_{\rm B}$, $I_{\rm A}$ and $I_{\rm F}$, corresponding to baseline, accretion and flaring components.  It is clear that there are different phases in the behaviour of \cyg, represented by the different branches of the colour-colour diagram. This would appear to represent the complex inter-dependency of more than one component of the X-ray emission. 

\begin{enumerate}

\vspace{5mm}
{\bf \item \noindent Component $I_{\rm B}$:}

As we have said earlier, $I_{\rm B}$ appears to describe the lowest possible X-ray intensity, or ``baseline'' as function of the rank number, \rank. The intensity is seen to fall to this level, but not below it, at both extremes of the Z-curve and possibly when the neutron star is momentarily obscured by intervening material. Component B smoothly softens between the observations on the Horizontal and Flaring Branches, as a result of a drop in intensity, predominantly in the hardest energy range. 

\vspace{5mm}
{\bf \item \noindent Component $I_{\rm A}$:}

The spectral shape of $I_{\rm A}$ remains constant while on the Horizontal Branch and also, with a different slope, on the Normal and Flaring Branches, only changing in flux while on either branch. The spectrum is seen to change between the Horizontal and Normal/Flaring Branches, when the sense of the correlation changes, with a harder spectrum on the Horizontal Branch. Like $I_{\rm B}$, this component is relatively constant on timescales of $10^{2-3}$~seconds. However, when the intensity drops into this region during non-correlated X-ray dips it does vary on these timescales. 

\vspace{5mm}
{\bf \item \noindent Component $I_{\rm F}$:}

$I_{\rm F}$ was defined such that it contains most of the $10^{1-2}$~seconds variability, as expressed by $\alpha(t)$ in Equation~\ref{intensityeqn}. The instantaneous X-ray intensity appears to change randomly within its envelope, except during X-ray bursts and non-correlated dips, which we therefore do not associate with variations in $I_{\rm F}$. The spectrum of this component shows a highly variable soft intensity, so that the total spectrum softens with increasing \rank. 
\end{enumerate}

$I_{\rm F}$, which represents the rapidly varying material is the softest of the three components and could be associated with material in a hot inner disc. At a given optical flux and hence inferred mass transfer rate, the flux in this component varies rapidly and appears to drive much of the rapid optical response, as can be seen by the correlated X-ray and optical flares in observations C and D. Such rapid variability could be associated with inhomogeneities in the accretion flow. 

$I_{\rm A}$, whose flux varies considerably, but with little change in the spectrum may be associated with the emission from the surface of the neutron star. Such variability could then be caused by the changing obscuration of the inner regions of the accretion flow by optically thick material. 

However, no complete physical interpretation of the decomposition is possible as yet. This is particularly evident when considering $I_{\rm B}$, which represents the lowest ('Baseline') level is the spectrum of the source when the accretion rate is at its lowest and highest levels. At these times there will be contributions from many regions within the binary, leading to a complicated spectrum. This component may include the non-thermal emission, which has been observed to increase with decreasing $\mdot$. Observations of similar variability of the so-called hard-tail in \cyg\ \cite{disalvo02} and other Z-sources would appear to support the existence of this variable hard component \cite{disalvo2000,frontera98}. In addition to this, it has been observed that the radio emission, which is linked to the mass outflow from the system, possibly in the form of a jet \cite{fender2000}, is greatest at the lowest \rank\ \cite{penninx88} and hence optical flux, as seen in our observations. 

The limited amount of data make it impossible to assess the importance of effects such as the binary phase and the long-term periodicity on $I_{\rm A}$ and indeed the other two components. However, it is intriguing that such a simple decomposition leads to three distinct components that are suggestive of the three regions thought to affect the X-ray emission in neutron star X-ray binaries.

%
%%%%%%%%%%%%%%%%%%%%%%%%%%%%%%%%%%%%%%%%%%%%%%%%%%%%%%%%%%%%%%%%%%%%%%%%%%%%%%%
%
\section*{Acknowledgments}
We thank John Cromer for writing, testing, and loading the software that allowed the LRIS CCD to read out continuously, and Bob Leach for helpful discussions.  We especially thank Tom Bida and Frederic Chaffee for their kindly letting us make changes to the LRIS system.  The optical data presented herein were obtained at the W.M. Keck Observatory, which is operated as a scientific partnership among the California Institute of Technology, the University of California and the National Aeronautics and Space Administration.  The Observatory was made possible by the generous financial support of the W.M. Keck Foundation. This research has made use of NASA's Astrophysics Data System Abstract Service.
%
%%%%%%%%%%%%%%%%%%%%%%%%%%%%%%%%%%%%%%%%%%%%%%%%%%%%%%%%%%%%%%%%%%%%%%%%%%%%%%%
%
\bibliographystyle{mn}
\bibliography{../../../Papers/ksobib}

\begin{thebibliography}{{{Kuulkers}, {van der Klis} \& {van Paradijs}}{1995}}

\bibitem[\protect\citefmt{{Bowyer} {\rm et~al.}}{1965}]{bowyer65}
{Bowyer}~S., {Byam}~E., {Chubb}~T., {Friedmann}~H., 1965, Science, 17, 894

\bibitem[\protect\citefmt{{Casares}, {Charles} \& {Kuulkers}}{1998}]{casares98}
{Casares}~J., {Charles}~P., {Kuulkers}~E., 1998, \apj, 493, L39

\bibitem[\protect\citefmt{{de Jong}, {van Paradijs} \&
  {Augusteijn}}{1996}]{dejong96}
{de Jong}~J.~A., {van Paradijs}~J., {Augusteijn}~T., 1996, \aa, 314, 484

\bibitem[\protect\citefmt{{Di Salvo} {\rm et~al.}}{2000}]{disalvo2000}
{Di Salvo}~T. {\rm et~al.}, 2000, \apjl, 544, L119

\bibitem[\protect\citefmt{{Di Salvo} {\rm et~al.}}{2002}]{disalvo02}
{Di Salvo}~T. {\rm et~al.}, 2002, \aa, 386, 535

\bibitem[\protect\citefmt{{Fender} \& {Hendry}}{2000}]{fender2000}
{Fender}~R.~P., {Hendry}~M.~A., 2000, \mnras, 317, 1

\bibitem[\protect\citefmt{{Frontera} {\rm et~al.}}{1998}]{frontera98}
{Frontera}~F. {\rm et~al.}, 1998, in The Active X-ray Sky: Results from
  BeppoSAX and RXTE.
\newblock p.~286+

\bibitem[\protect\citefmt{{Giacconi} {\rm et~al.}}{1973}]{giacconi73}
{Giacconi}~R., {Gurksy}~H., {Kellogg}~E., {Levinson}~R., {Schreier}~E.,
  {Tananbaum}~H., 1973, \apj, 184, 227

\bibitem[\protect\citefmt{{Hasinger} \& {van der Klis}}{1989}]{hasinger89}
{Hasinger}~G., {van der Klis}~M., 1989, \aa, 225, 79

\bibitem[\protect\citefmt{{Hasinger} {\rm et~al.}}{1990}]{hasinger90}
{Hasinger}~G., van~der {Klis}~M., {Ebisawa}~K., {Dotani}~T., {Mitsuda}~K.,
  1990, \aa, 235, 131

\bibitem[\protect\citefmt{{Hjellming} {\rm et~al.}}{1990}]{hjellming90}
{Hjellming}~R.~M., {Han}~X.~H., {Cordova}~F.~A., {Hasinger}~G., 1990, \aa, 235,
  147

\bibitem[\protect\citefmt{{Kahn} \& {Grindlay}}{1984}]{Kahn84}
{Kahn}~S.~M., {Grindlay}~J.~E., 1984, \apj, 281, 826

\bibitem[\protect\citefmt{{Kuulkers} {\rm et~al.}}{1997}]{kuulkers97b}
{Kuulkers}~E., {van der Klis}~M., {Oosterbroek}~T., {Van Paradijs}~J.,
  {Lewin}~W. H.~G., 1997, \mnras, 287, 495

\bibitem[\protect\citefmt{{Kuulkers}, {van der Klis} \& {van
  Paradijs}}{1995}]{kuulkers95}
{Kuulkers}~E., {van der Klis}~M., {van Paradijs}~J., 1995, \apj, 450, 748

\bibitem[\protect\citefmt{{Kuulkers}, {van der Klis} \&
  {Vaughan}}{1996}]{kuulkers96}
{Kuulkers}~E., {van der Klis}~M., {Vaughan}~B., 1996, \aa, 311, 197

\bibitem[\protect\citefmt{{Oke} {\rm et~al.}}{1995}]{oke95}
{Oke}~J.~B. {\rm et~al.}, 1995, \pasp, 107, 375

\bibitem[\protect\citefmt{{Oke}}{1990}]{oke90}
{Oke}~J.~B., 1990, \aj, 99, 1621

\bibitem[\protect\citefmt{{Orosz} \& {Kuulkers}}{1999}]{orosz99}
{Orosz}~J.~A., {Kuulkers}~E., 1999, \mnras, 305, 132

\bibitem[\protect\citefmt{{Penninx} {\rm et~al.}}{1988}]{penninx88}
{Penninx}~W., {Lewin}~W. H.~G., {Zijlstra}~A.~A., {Mitsuda}~K., {van
  Paradijs}~J., 1988, \na, 336, 146

\bibitem[\protect\citefmt{{Smale}}{1998}]{smale98}
{Smale}~A.~P., 1998, \apjl, 498, L141

\bibitem[\protect\citefmt{{van der Klis}}{2001}]{vdk2001}
{van der Klis}~M., 2001, \apj, 561, 943

\bibitem[\protect\citefmt{{van Paradijs} {\rm et~al.}}{1990}]{paradijs90}
{van Paradijs}~J. {\rm et~al.}, 1990, \aa, 235, 156

\bibitem[\protect\citefmt{{Vrtilek} {\rm et~al.}}{1990}]{vrtilek90}
{Vrtilek}~S., {Raymond}~J., {Garcia}~M., {Verbunt}~F., {Hasinger}~G.,
  {Kurster}~M., 1990, \aa, 235, 162

\bibitem[\protect\citefmt{{Wijnands} \& {van der Klis}}{2001}]{wijnands01}
{Wijnands}~R., {van der Klis}~M., 2001, \mnras, 321, 537

\bibitem[\protect\citefmt{{Wijnands}, {Kuulkers} \& {Smale}}{1996}]{wijnands96}
{Wijnands}~R. A.~D., {Kuulkers}~E., {Smale}~A.~P., 1996, \apjl, 473, L45

\end{thebibliography}
%
%%%%%%%%%%%%%%%%%%%%%%%%%%%%%%%%%%%%%%%%%%%%%%%%%%%%%%%%%%%%%%%%%%%%%%%%%%%%%%%
\label{lastpage}
\end{document}